\documentclass[twoside,english,nofootinbib,preprintnumbers,aps,11pt]{revtex4}

\usepackage{euscript,amssymb}
\usepackage{amsthm,latexsym}
\usepackage{graphicx}
\usepackage{amsfonts}
\usepackage{amsmath}
\usepackage{amssymb}
\usepackage{fancyhdr}
\usepackage{epsfig}
\usepackage{color}
\usepackage{esint}
\usepackage{soul}
\usepackage{afterpage}
\usepackage[papersize={8.5in,11in}]{geometry}

\usepackage[normalem]{ulem}
\usepackage{color}

\newcommand{\be}{\begin{equation}}
\newcommand{\ee}{\end{equation}}

\begin{document}


\title{Near Horizon Geometries and Black Hole Holograph}


\author{Jerzy Lewandowski$^1$}

\email[]{Jerzy.Lewandowski@fuw.edu.pl}

\author{Istvan Racz$^2$}

\email[]{racz.istvan@wigner.mta.hu}

\author{Adam Szereszewski$^1$}

\email[]{Adam.Szereszewski@fuw.edu.pl}

 \affiliation{\vspace{6pt} $^1$Faculty of Physics, University of Warsaw, Pasteura 5, 02-093 Warsaw, Poland}

\affiliation{\vspace{6pt} $^2$Wigner RCP, Konkoly Thege Mikl\'os \'ut 29-33, H-1121 Budapest, Hungary}

\begin{abstract}   

Two quasi local approaches  to black holes are combined:  Near Horizon Geometries (NHG) and 
stationary Black Hole Holographs (BHH). Necessary and sufficient   conditions on  BHH data  
for the emergence of  NHGs as resulting vacuum solutions to Einstein's equations are found.       

\end{abstract}

\date{\today}

\pacs{???}

\maketitle

\subsubsection{Introduction}

In this letter we combine results of two topics  of the quasi local theory of  black holes (BH).  The first one is  the theory of 
near horizon geometries (NHG) of  extremal BHs \cite{Horowitz,LivRevNHG,Reall}. They are exact solutions to Einstein's equations  obtained by 
a naturally defined limit of  neighborhoods of extremal (degenerate) Killing horizons.  The first examples were derived from  
the extremal Reissner-Nordstr\"om solution and from the extremal Kerr.  A larger family of examples  (defined 
modulo an equation, that has to be solved, though) is set  by the Kundt's class of solutions to Einstein's equations 
\cite{PLJ,Podolsky1,Podolsky2,Podolsky3,AszerLewWal1,AszerLewWal2,ExactSolutions}.  The second topic is the recent stationary Black Hole holograph (BHH) \cite{Racz1,Racz2}. This approach relies on the characteristic Cauchy problem for the electrovacuum Einstein's equations.  If the  transversal to each other null surfaces are non-expanding, then  they become components of a bifurcated Killing horizon.   The motivation for the current paper is an observation, that the NHGs also admit bifurcated Killing horizons.  That makes them a special case of the BHHs. In the current letter we present a solution to the inverse problem, namely, we find conditions on the BHH data that are necessary and sufficient for the corresponding hologram spacetime to be a NHG.  Our result may be considered as the first step in using  the BHH construction  in a quest for an interesting generalization of the idea of  NHG. For simplicity, we will restrict  here to 4d spacetimes and the vacuum Einstein's equations.                   

A BHH data $(S,g,\omega)$ is:  a compact 2-manifold $S$ (a BHH space) endowed with  a metric tensor 
(a BHH metric tensor)
\begin{equation} 
g\ =\ g_{AB}\,dx^A dx^B
\end{equation}
and a 1-form (a BHH 1-form)
\begin{equation} 
	\omega\ =\ \omega_A \,dx^A\
\end{equation}
where $(x^A)=(x^1,x^2)$ is a local coordinate system at $S$. Note that this  is a geometric version  of the original definition  \cite{Racz1,Racz2}.
The corresponding  hologram is a 4d spacetime in which  the 2-space  $S$  becomes the intersection  between two  non-expanding 
null surfaces (Non-Expanding Horizons \cite{LivRevIH,LPhigh,ABL}), while $g$ becomes the metric tensor induced in $S$.   
The 1-form $\omega$ becomes the pullback to $S$ of the rotation 1-form potential of one of the horizons, and, respectively,  
minus the rotation 1-form potential of the other one pulled back to $S$. The spacetime geometry is determined via the characteristic 
Cauchy problem for  vacuum Einstein's equations in the causal future and in the past of the intersection $S$.  The BHH theorem states  
that in this spacetime the non-expanding horizons set a bifurcated Killing horizon.  Among all the black hole spacetimes obtained in 
that way there are also all the NHGs. Indeed, it is known, that each NHG contains a bifurcated Killing horizon \cite{rw,PLJ}.  We will find 
below necessary and sufficient conditions on $(S,g,\omega)$  for the corresponding hologram to be a NHG.

\subsubsection{The Black Hole Holograph}

Given a BHH data $(S,g,\omega)$  the hologram spacetime  manifold $M$  
has the product topology
\begin{equation}
M\ \sim\ S\times\mathbb{R}\times\mathbb{R}\,.
\end{equation}
The coordinates $(x^A)$  defined on  $S$, as well as coordinates $u$ and $v$ defined on the first, and the second factor $\mathbb{R}$,
respectively, are naturally extended to the Cartesian product. The surfaces
\begin{equation} 
	N_1\ \ \ {\rm such\ that} \ \ \ u\ = 0\ \ \ \ \  {\rm and} \ \ \ \ \ N_2\ \ \ {\rm such\ that}\ \ \    v\ =\ 0 ,
\end{equation}
respectively, are assumed to be null and  non-expanding with respect to the resulting spacetime geometry,
while $S$ is identified with the surface $u = v = 0$ in $M$.  
According to the standard characteristic Cauchy problem for vacuum Einstein's equations, in the smooth case, the spacetime geometry 
is determined  (up to remaining diffeomeorphisms)  
in the wedges $u\ge 0, v\ge 0$ and  
$u \le 0, v\le 0$, in some neighborhood of $S=N_1 \cap N_2$, 
provided that the following conditions
hold at the surfaces  $N_1$ and $N_2$, and at $S$, respectively:
 \begin{itemize}
   \item The pullback of the spacetime metric to each of the surfaces $N_1$, and $N_2$ respectively,  is the following degenerate metric
      \begin{equation} \label{hol1}  
          g_{AB}\,dx^Adx^B.   
      \end{equation}
   \item The vectors  $\ell = \partial_u$, $n=\partial_v$ are future oriented and satisfy
      \begin{equation} \label{hol2}
         \nabla_{\ell}\ell |_{N_1}\ =\ 0\,, \ \ \ \ \ \ \ \nabla_n n|_{N_2}\ =\ 0\,. 
      \end{equation}
   \item The pull back to $S$ of the 1-form $-n_\mu\nabla_\nu\,\ell^\mu$ is 
       \begin{equation}\label{hol3}
          -n_\mu\nabla_A \ell^\mu |_S\ =\ \omega_A\,.
       \end{equation} 
\end{itemize}

The BHH theorem \cite{Racz1,Racz2} states, that the spacetime metric tensor  determined by the data  
admits a Killing vector $K$, that, using the remaining diffeomorphisms,  can be given the form
 \begin{equation}
K\ =\ u\,\partial_u \ -\ v\,\partial_v \,.
\end{equation} 
Therefore, the surfaces $N_1$ and $N_2$ form a (non-extremal)  bifurcated Killing horizon
while the pullback of the 1-form $\omega_\nu=-n_\mu\nabla_\nu\,\ell^\mu$ to $N_2$ is its  rotation 1-form potential.
In this sense the construction works as a stationary BH holograph: given  any 2-dimensional 
data $(S,g,\omega)$ it produces  4-dimensional spacetime in 
the domain of dependence of the bifurcate Killing horizon, $N_1 \cup N_2 $.

\subsubsection{NHG from BHH}

Suppose now, that a  BHH data $(S,g,\omega)$ satisfies the following equation
\begin{equation} \label{theeq} 
\omega_{(A;B)} \ +\ \omega_A\omega_B\ -\frac{1}{2}R_{AB}\ =\ 0    
\end{equation}
where by `$;$' we denote the torsion free covariant derivative defined on $S$ by the metric $g$,
and $R_{AB}$ is the Ricci tensor of $g$.  This equation is soluble \cite{PLJ,Jez,JezKerr} only when $S$ is either a topological 2-sphere 
$$ S\ =\ S_2 $$
or a 2-torus
$$ S\ =\ S_1\times S_1 \,.$$
In the latter case, the only solution is 
$$ \omega_A=0=R_{AB} \,,$$
therefore we will be assuming henceforth, that the manifold $S$ is a 2-sphere  $S_2$. Whenever (\ref{theeq}) holds, 
the hologram metric tensor can be written down explicitly. Indeed,  the following  metric tensor 
\begin{equation}\label{NHG}    
ds^2\ =\  -2\,du\Big(dv -2\,v\,\omega -\frac{1}{2}\,v^2\left[\omega_{A;}{}^A +2\omega_A\omega^A\right]du\Big) \ +\ g_{AB}\,dx^Adx^B   
\end{equation}
is an exact solution of the vacuum Einstein equations  \cite{PLJ} that matches the hologram data (\ref{hol1})-(\ref{hol3}). Owing to uniqueness (mod diffeomorphisms) in BHH \cite{Racz2}, this is the corresponding BH hologram.   
Such geometries are called Near Horizon.   A remarkable property of this BH hologram  (\ref{NHG}) is  emergence of a second  Killing vector field, namely
\begin{equation} 
L\ =\ \partial_u \,.  
\end{equation} 
The surface $N_2$ is extremal Killing horizon of the Killing vector field $L$, still being a component of the bifurcated non-extremal 
horizon of the Killing vector field $K$.  Therefore, our first conclusion is,  that  every  BHH  data $(S,g,\omega)$  such that  the 
equation  (\ref{theeq}) is satisfied,  defines a NHG with the extremal Killing horizon $N_2$.

\subsubsection{Flipped NHG BHH data}  

The gauge freedom we have in setting up the initial data for BHH \cite{Racz2} yields some ambiguity in identifying NHGs. For instance, 
 condition (\ref{theeq}) is not necessary as other BHH data may also define a NHG as the hologram spacetime.  
For example, the following transformation in the space of the holographic data 
\begin{equation}   
(S,g,\omega)\  \mapsto\  (S,g,-\omega)  
\end{equation}
corresponds to switching  of the factors in $S\times\mathbb{R}\times\mathbb{R}$, namely
\begin{equation}   
(x^A,u,v)\  \mapsto\  (x^A,v,u) \,, 
\end{equation}
because on $S={N_1\cap N_2}$
$$  -(\partial_v)_\mu\nabla_A (\partial_u)^\mu\ =\   (\partial_u)_\mu\nabla_A (\partial_v)^\mu $$
holds.
Hence, every data $(S,g,\omega)$ which satisfies the switched equation (\ref{theeq}), that is
\begin{equation} \label{theeq'} \omega_{(A;B)} \ -\ \omega_A\omega_B\ +\frac{1}{2}\,R_{AB}\ =\ 0    \end{equation}  
also defines a NHG, this time with   the extremal horizon $N_1$.

\subsubsection{A general case of NHG  from BHH}  

An analogous gauge freedom explains that the most general form of condition (\ref{theeq})  upon 
which    BHH data  still gives rise  to   NHG such that $N_2$ becomes the extremal Killing horizon comes with the following gauge transformation 
\begin{equation}   
(S,g,\omega)\  \mapsto\  (S,g,\omega + d\lambda) \end{equation}
where   $\lambda:S\rightarrow\mathbb{R}$ is an arbitrary function (differentiable suitable number of times).
It may be obtained by  the following coordinate transformation 
\begin{equation}   
(x^A,u,v)\  \mapsto\  (x^A,e^{-\lambda} u,e^{\lambda} v) \,. 
\end{equation}
It follows that every data $(S,g,\omega)$ such that there is a function $\lambda:S\rightarrow S$ such that
\begin{equation} \label{theeq''} 
\omega_{(A;B)}+\lambda_{;AB} \ +\ (\omega_A+\lambda_{,A})(\omega_B+\lambda_{,B})\ -\frac{1}{2}R_{AB}\ =\ 0    
\end{equation}
also defines a NHG. This non-linear equation on the unknown function $\lambda$ can be written as a linear equation on a new 
nowhere vanishing function 
$$f\ :=\ e^\lambda. $$
Indeed, that substitution turns the equation (\ref{theeq''})  into  (below, $`\ D\ '\ \equiv\ `\ ;\ '$)
\begin{equation} \label{theeq'''}  \left( D_AD_B \ +\ \omega_A D_B+ \omega_B D_A  + \omega_{(A;B)} \ +\ 
\omega_A\omega_B\ -\frac{1}{2}R_{AB}\right)f\ =\ 0  \,.
\end{equation}

\subsubsection{Necessary and sufficient condition on BHH to give rise to NHG with respect to $N_2$.}

It turns out, that not every BHH data $(S_2,g,\omega)$ admits a solution $f$ to (\ref{theeq'''}). 
The solubility  conditions are expressed by the following BHH data invariants  \cite{ABL,LPuniq}: 
   \begin{description}
     \item[$(i)$] the Ricci scalar 
       \begin{equation} R\ =\ R_{AB}\,g^{AB} \end{equation}
 of the 2d metric tensor $g$, and
   \item[$(ii)$] the rotation invariant 
  $$ \star d\omega  $$ 
 and its scalar potential $U$, defined by the following equation
\begin{equation}  
 \Delta U\ =\ \star d\omega\,,  
\end{equation} 
where $\Delta$ and $\star$ are the Laplace operator and Hodge star respectively defined 
on $S_2$ by the metric $g$. 
\end{description}
 The necessary solubility conditions are

\begin{enumerate}
   \item  The complex valued function  $R + 2i\Delta U$ nowhere vanishes, i.e. 
      \begin{equation}\label{sol1}   
              R + 2i\Delta U\ \not=0\ \ \ {\rm for\ \ every}\ \ \ x\in S_2\,. 
      \end{equation}
   \item
      \begin{equation}\label{sol2}   
             \frac{R + 2i\Delta U}{R - 2i\Delta U}e^{6iU}\ =\  {\rm const} \,.
      \end{equation}
   \item The function 
      \begin{equation}\label{f'} 
                   \tilde{f} \ :=\ \left(R - 2i\Delta U\right)^{-\frac{1}{3}}e^{iU} 
      \end{equation}
and the 1-form 
      \begin{equation} \label{omega'}
                   \tilde{\omega} \ :=\ \star dU . 
      \end{equation}
satisfy equation (\ref{theeq'''}), that is
      \begin{equation} \label{theeq''''}  
	\left( D_AD_B \ +\ \tilde{\omega}_A D_B+ \tilde{\omega}_B D_A  + \tilde{\omega}_{(A;B)} \ +\ 
          \tilde{\omega}_A\tilde{\omega}_B\ -\frac{1}{2}R_{AB}\right)\tilde{f} =\ 0  \,.
     \end{equation}
\end{enumerate}

The meaning of Condition 2  is, that the function $\left(R + 2i\Delta U\right)e^{3iU}$ is real valued 
up to a constant factor. The factor can be absorbed into the potential $U$.  

Notably Conditions 1--3 are also sufficient for the 
existence of a nowhere vanishing function $f$ satisfying equation (\ref{theeq'''}).
To see this note that such an $f$ can be given as
   \begin{equation} f\ :=\ B\,\tilde{f}\,, \end{equation}
where $\tilde{f}$ is a solution to (\ref{f'}), whereas the 
function $B$ is given as
\begin{equation} 
    \Delta {\rm ln}B\ :=\  \star d\star\omega \,. 
\end{equation}

\subsubsection{Non-rotating BHH data}

An example of a BHH data $(S_2,g,\omega)$, such that our Conditions 1--3 are not satisfied is 
a non-rotating case   when 
\begin{equation} \label{omega'0}
    d\omega \ =\ 0 \,.
\end{equation}
In this case  the equations (\ref{f'})-(\ref{theeq''''})  imply \cite{LPuniq,ABL}
$$ 0\ =\ \int_{S_2} d^2x \sqrt{{\rm det} \,g}\left(\Delta - \frac{1}{2}R\right)R^{-\frac{1}{3}} \ =\  
- \frac{1}{2}\int_{S_2} d^2x \sqrt{{\rm det} \, g} \left(R^\frac{1}{3}\right)^2 ,    $$
a condition requiring the vanishing of $R$ throughout $S_2$.
This, however, leads to a contradiction as a 2-sphere does not admit a flat metric tensor
which, in turn, verifies non-existence of non-rotating vacuum NHG \cite{ChruscielReallTod}.

\subsubsection{Summary and outlook}

The subset of  BHH data that corresponds to NHGs was identified. 
It was shown that, up to the flips of the horizons, a BHH data
$(S_2,g,\omega)$ gives rise to a NHG if and only if 
Conditions 1--3 hold.  A general exact solution to the involved constraints is not known. 
The found relation between the BHH construction and the NHGs may 
play significant role in attempting to give suitable generalizations of the 
concept of NHG. The NHGs are exact solutions to Einstein's
equations that at the same time provide the 0th order in a suitable expansion of spacetime metric about an extremal
Killing horizon. 
In the non-extremal case a suitable  generalization of the NHGs is not known yet.

\subsubsection{Acknowledgements}

This work was partially supported by the Polish National
Science Centre\\ grant No. 2015/17/B/ST2/02871 and by the Hungarian NKFIH grant K-115434.

\end{document}